\documentclass[final]{ias2}
\usepackage{graphicx} 
\usepackage{multirow}
\usepackage{array} 

\usepackage{hyperref} 
\newcommand\beq{\begin{equation}}
\newcommand\eeq{\end{equation}}
\newcommand\bea{\begin{eqnarray}}
\newcommand\eea{\end{eqnarray}}

\begin{document}
\markboth{Pramana class file for \LaTeX 2e}{Murthy M V N, et. al.}

\title{A phenomenological approach to the equation of state of a unitary 
Fermi gas}

\author[mur]{M.V.N. Murthy}
\email{murthy@imsc.res.in}

\author[bra]{M. Brack}
\email{matthias.brack@physik.uni-regensburg.de}

\author[bha]{R.K. Bhaduri}
\email{bhaduri@physics.mcmaster.ca}

\address[mur]{The Institute of Mathematical Sciences, Chennai 600113, 
India.}

\address[bra]{Institute for Theoretical Physics, University of Regensburg,
Regensburg, Germany}

\address[bha]{Department of Physics, McMaster University, 
Hamilton, Ont. L8S4M1, Canada}

\begin{abstract} 

We propose a phenomenological approach for the equation of state of a 
unitary Fermi gas. The universal equation of state is parametrised in 
terms of Fermi-Dirac integrals. This reproduces the experimental data 
over the accessible range of fugacity and normalised temperature, but 
cannot describe the superfluid phase transition found in the MIT 
experiment \cite{ku}. The most sensitive data for compressibility and 
specific heat at phase transition can, however, be fitted by introducing 
into the grand partition function a pair of complex conjugate zeros 
lying in the complex fugacity plane slightly off the real axis.

\end{abstract}

\keywords{unitary Fermi gas, equation of state}
\pacs{05.30.Fk  64.10.+h }
\maketitle

\section{Introduction}

Recently, the thermodynamics of a unitary gas of fermionic atoms has been 
in the focus of experimental investigations \cite{1,2,hori,ku}. 
In a unitary gas, the inter-atomic 
interaction between neutral fermionic atoms is adjusted using the Feshbach 
resonance \cite{fr}, so that the scattering length goes to $\pm \infty$. 
Such a gas has properties that are universal or scale independent 
\cite{ho}. The experimental confirmation of the universal nature of the 
equation of state (EOS) of a gas of neutral fermionic atoms has 
therefore given fresh impetus to its theoretical understanding 
\cite{3,3p}. In a recent paper \cite{bvm}, an ansatz for the grand 
potential of a spin balanced two-component fermion gas was introduced 
through a virial expansion in powers of the fugacity variable $z$. This 
ansatz for the interaction part of the virial coefficients could fit the 
experimental data up to about $z=7$, surprising in view of the fact that 
it was meant to be a high temperature expansion for small $z$. For $z>7$, 
i.e., at low temperatures, the virial expansion was found 
to become unphysical.

In this paper we propose a novel phenomenological approach to describe 
the EOS, that agrees with experimental data all the way to very low 
temperatures and reproduces some of the zero temperature properties 
quantitatively. Following Sommerfeld \cite{sfeld}, it would seem that at 
low temperatures when $z\gg 1$, $x=\ln z$ is a more suitable expansion 
parameter. Since a unitary gas introduces no extra length scales than 
already present in the ideal gas, we go one step further and express the 
grand potential in terms of a simple combination of two Fermi-Dirac 
integrals \cite{pathria}. This allows us to fit the experimental data 
quite accurately, and at the same time to reproduce the correct second 
virial coefficient at small $z$, i.e., for high temperatures.

Furthermore, experimentally, a phase transition to super-fluidity is 
observed around $T/T_F\simeq 0.16$, evidenced by peaks in the heat 
capacity and compressibility \cite{ku}. To reproduce these features in 
our model, we introduce a phenomenological term in the grand partition 
function, which in the zero-width limit yields a singularity 
in the free energy and hence describes a phase transition.
       
In Sec.\ II, we first introduce our new phenomenological ansatz 
and show that it reproduces the universal function $h(z)$, which is the 
ratio of pressures of the spin-balanced two-component unitary gas and 
free Fermi gas, over a large range of experimentally available 
fugacities $z$. In Sec.\ III, we  compute all the thermodynamic 
quantities for which experimental data are available and show that our 
phenomenological ansatz indeed incorporates the essential features of 
the data.

\section{The phenomenological equation of state}

The grand potential $\Omega$ of the unitary gas is related to the grand 
partition function ${\cal Z}$ by the thermodynamical relation
\begin{equation}
\Omega=-PV=-k_BT\ln {\cal Z}\,,
\end{equation}
where $P$ and $V$ are pressure and volume, respectively, $T$ is the 
temperature, and $k_B$ the Boltzmann constant. The grand partition function
is defined by
\begin{equation}
{\cal Z}(\beta,z)=\sum_{N=0}^\infty Z_N(\beta)\,z^N, \qquad 
                  \beta = 1/k_BT\,,
\end{equation}
where $Z_N(\beta)$ is the canonical $N$ particle partition function.
Note that in the above series, the dependences on $\beta$ and $z$ are
mixed. However, for the ideal free Fermi gas, ${\cal Z}$ has the form
\begin{equation}
\ln {\cal Z}_F(\beta,z)=\frac{V}{\lambda^3}\,2f_{5/2}(z)\,,
   \quad \lambda = \left(\frac{2\pi\hbar^2\beta}{m}\right)^{\!\!1/2}\!,
\label{Zfree}
\end{equation}
in which the $z$ dependence has separated out and is entirely coming through 
the function $f_{5/2}(z)$ which is one of the Fermi-Dirac integrals defined 
\cite{pathria} as
\begin{equation}
f_\nu(x)=\frac{1}{\Gamma(\nu)}\int_0^\infty \frac{y^{(\nu-1)}dy}
{1+e^{(y-x)}}\,.
\label{eqn2}
\end{equation}

For a unitary gas, a similar separation of variables $\beta$ and $z$ also 
takes place \cite{ho}. We therefore define a universal function $F_P(z)$ by
\begin{equation}
\ln {\cal Z}(\beta,z)= (V/\lambda^3)\,F_P(z)\,.
\label{fP}
\end{equation}
In terms of this function, we define the universal thermodynamic function 
$h(z)$ by
\begin{equation}
h(x)=\frac{\Omega}{\Omega_F}=\frac{P}{P_F}
    =\frac{F_P(x)}{2f_{5/2}(x)}\,, \qquad x=\ln(z)\,,
\label{eqn1}
\end{equation}
where $\Omega_F$ and $P_F$ are the grand potential and pressure of the 
untrapped ideal Fermi gas, respectively. Note that in this quantity the
dependence on temperature and length scales drops out, so that it is
universal and scale independent (see also \cite{1}).

%\newpage

The all-important function $F_P(x)$ encodes the thermodynamic properties 
of the unitary gas of fermionic atoms. We make the important assumption 
that $F_P(x)$ can be written as a linear superposition of Fermi-Dirac 
integrals since this ensures universality.  Thus we introduce  
the function $F_P(x)$ through the following phenomenological ansatz:
\begin{equation}                                                   
F_P(x)=2[f_{5/2}(x)+4(f_{5/2}(x)-f_{3/2}(x))]+g(x)\,,
\label{eqn3}
\end{equation}   
where the factor 2 in front accounts for spin degeneracy. 
This ansatz is further guided by the following considerations: 

1. The function $h(x)$ obeys universality, i.e., it depends only on the 
fugacity $z=\exp(x)$, but not on any length scale or other system variable.

2. The leading term $f_{5/2}(x)$ in Eq.\ (\ref{eqn3}) is simply 
that of the free non-interacting Fermi gas given in Eq.\ (\ref{Zfree}).

3. The second term $4(f_{5/2}(x)-f_{3/2}(x))$ describes the 
contribution from the interactions. By definition this term does not 
contribute to the linear term in $z$ in the high temperature expansion 
of $F_P(x)$ (cf.\ \cite{pathria}). Furthermore, the linear superposition 
of Fermi-Dirac integrals is determined to yield the exact interaction
part of the second virial coefficient $\Delta b_2$ \cite{HoMu} in the 
high temperature limit. This choice, however, does not yield the correct 
third virial coefficient that is known to great accuracy, nor the estimated 
fourth virial coefficient \cite{liu,rakshit}. Nevertheless, the high 
temperature properties that we obtain  
still give excellent agreement with experimental results. On the other 
hand, the zero temperature properties are entirely determined by the 
function $f_{5/2}(x)$. The overall factor 5 of $f_{5/2}(x)$ 
allows for a good description also of the zero temperature properties of 
the unitary gas (see the detailed discussion to Fig.\ \ref{fig2} below). 

4. The function $g(x)$ in Eq.\ (\ref{eqn3}), which is implicitly a function 
of the fugacity $z=\exp(x)$, is introduced in order to describe the phase
transition that has been observed in the experimental data \cite{ku}.
For this purpose, we write the grand partition function ${\cal Z}$ in the 
complex $z$ plane as
\begin{equation}
{\cal Z}={\widetilde {\cal Z}}\left[\left(1-\frac{z}{z_c+i\epsilon}\right)\!
         \left(1-\frac{z}{z_c-i\epsilon}\right)\right]^{(V/\lambda^3)}, 
\label{pinch}
\end{equation}
with real $z_c$ and $\epsilon$, where ${\widetilde {\cal Z}}$ describes the 
system without phase transition. At $z=z_c\pm i\epsilon$, this function goes
to zero, causing a logarithmic singularity of the free energy in the limit 
$\epsilon\to 0$. 
The power $(V/\lambda^3)$ of the zeros is required to preserve the universality 
of $h(x)$. For the function $g(z)$, which is found through Eqs.\ (\ref{eqn3}) 
and (\ref{fP}), this yields
\begin{equation}
g(z) = \ln\left[\frac{(z_c-z)^2+\epsilon^2}{z_c^2+\epsilon^2}\right]
       +2\,\frac{z}{z_c}\,.
\label{eqn4}
\end{equation}
The last term in Eq.\ (\ref{eqn4}) is introduced such that $g(x)$ gives no 
contribution to the first-order virial coefficient in the high temperature
(i.e., small-$z$) expansion of $F_P(x)$. The choice of $z_c$ and $\epsilon$ is 
guided by a fit to the experimental data on compressibility and specific heat: 
while $z_c$ is given by the critical temperature of the phase transition, 
$\epsilon$ is governed by the width of the transition region. 

We find that $z_c=13.5$ and $\epsilon=6.3$ give the best fits to the MIT data 
for compressibility and specific heat, seen in Fig.\ \ref{fig3} below. While 
the form (\ref{eqn4}) of $g(x)$ works well throughout the phase transition 
region and all the way to small $z$ (i.e., to high temperatures), it becomes 
unphysical in the zero $T$ limit where we are forced to use the ansatz 
(\ref{eqn3}) with $g=0$. 
\begin{figure}[ht]
\begin{center}
\includegraphics[width=1.0\columnwidth]{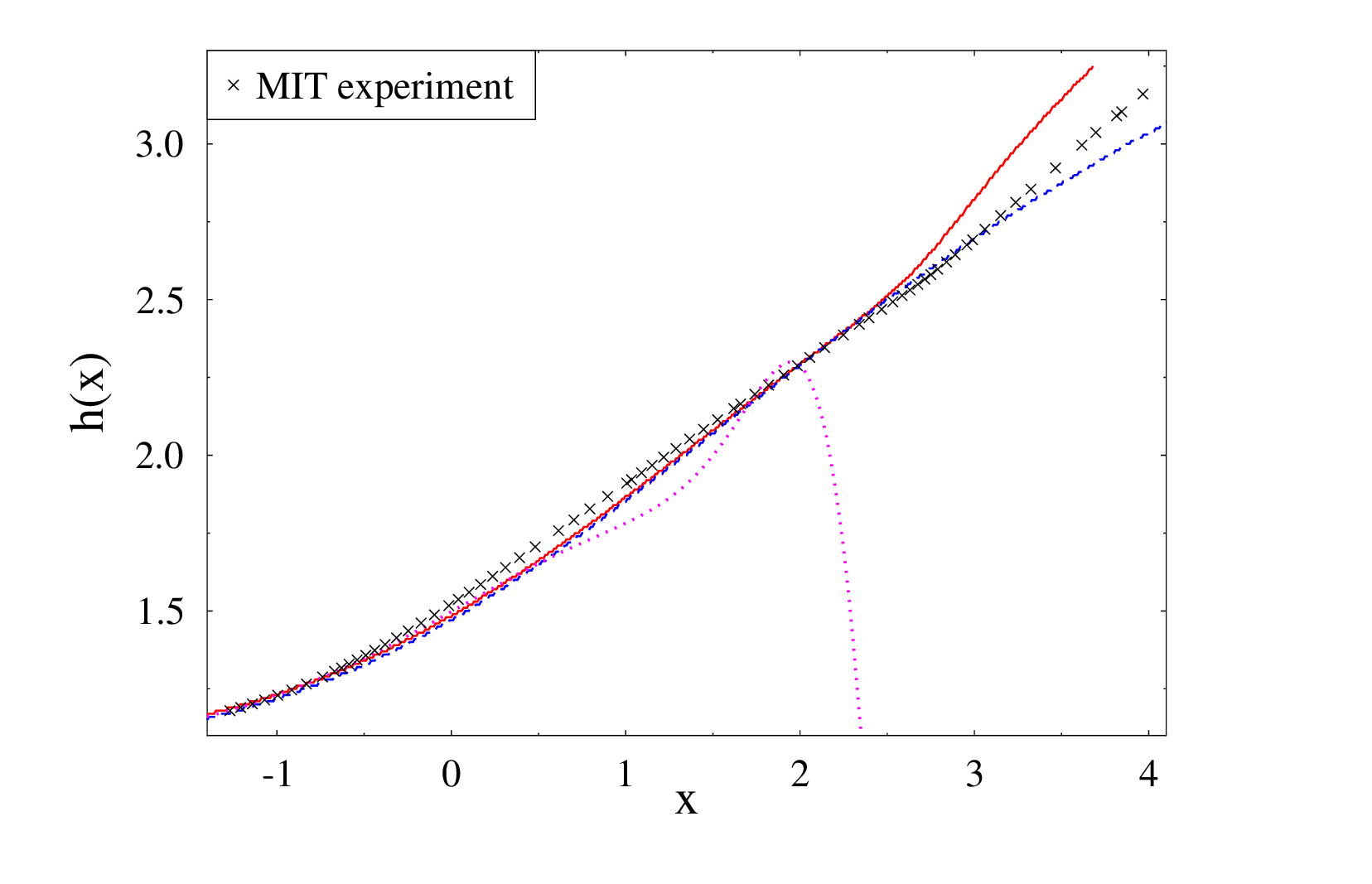}
\caption{(Colour online) The function $h(x)$ for the untrapped unitary 
Fermi gas as a function of $x=\ln(z)$. The crosses represent the 
experimental MIT data by Ku {\it et al} \protect\cite{ku}. Our result 
including the full $g(x)$ in (\ref{eqn3}) is shown by the (red) solid 
line. The result obtained by putting $g(x)=0$ is shown by the (blue) 
dashed line. We also show the results for the virial ansatz 
\protect\cite{bvm} by the (magenta) dotted line.} 
\label{fig1} 
\end{center}
\end{figure} 

In Fig.\ \ref{fig1} we compare our numerical results with the experimental 
data of the MIT group \cite{ku} for the universal function $h(x)$ given in 
Eq.\ (\ref{eqn1}). The solid line corresponds to the full expression 
(\ref{eqn3}) for $F_P(x)$, while the dotted line is obtained using $g(x)=0$.
We see that the our phenomenological ansatz closely follows the data up to
$x \sim 2.5$, while the high temperature virial ansatz introduced in 
\cite{bvm} fails much earlier. For large $x$, the results obtained with 
$g(x)=0$ and with $g(x)$ given in (\ref{eqn4}) lie on either side of the 
data, the solid line showing that $g(x)$ becomes unphysical for $x>3$. In 
all calculations presented henceforth, we have put it to zero for 
$z>z_{max}=27$, corresponding to $x_{max}=3.3$ and a temperature $T_{min}
=0.1\, T_F$, below which there are essentially no data points found in the 
figures below.

\section{Thermodynamical properties}

Encouraged by the good agreement over a large range of $x$ for 
our universal function $h(x)$, we now consider the calculation of basic
thermodynamic observables. Following Ku {\it et al.} 
\cite{ku}, we write for the normalised pressure
\begin{equation}
\widetilde{p}=\frac{P}{P_0} =\frac{5T}{2T_F}\frac{F_P(x)}{F'_P(x)}
=\frac{5}{3}\frac{E}{NE_F}\,,
\label{eqn5}
\end{equation}
where $P$ and $E$ are pressure and energy, respectively, of the interacting 
gas. The quantities used for the normalisation in the denominators above
are all evaluated for the non-interacting gas: $P_0$ is the pressure at zero 
temperature, $T_F$ the Fermi temperature, and $E_F$ the Fermi energy; the
latter two are related by $E_F=k_B T_F$.
The prime here and below denotes derivative with respect to $x$. 
The normalised temperature is given by
\begin{equation}
\frac{T}{T_F}=\frac{k_BT}{E_F}=\frac{4\pi}{[3\pi^2F'_P(x)]^{2/3}}\, .
\label{eqn6}
\end{equation}
The entropy, also related to pressure, is given by
\begin{equation}
\frac{S}{Nk_B}=\frac{T_F}{T}\left(\widetilde{p}-\frac{\mu}{E_F}\right)=
\frac{5F_P(x)}{2F'_P(x)}-\ln(z)\,.
\label{eqn7}
\end{equation}
The chemical potential $\mu$, normalised with respect to the non-interacting
Fermi energy $E_F$, is given by
\begin{equation}
\frac{\mu}{E_F}=\widetilde{p} -\frac{TS}{T_F N k_B}\,.
\label{eqn8}
\end{equation}
Analogously, the normalised compressibility is given by
\begin{equation}
\widetilde{\kappa}=\frac{\kappa}{\kappa_0} 
=\frac{2T_F}{3T}\frac{F''_P(x)}{F'_P(x)}\,.
\label{eqn9}
\end{equation}
The specific heat at constant volume is given by
\begin{equation}
\frac{C_V}{Nk_B} =
\frac{15}{4}\frac{F_P(x)}{F'_P(x)}
-\frac{9F'_P(x)}{4F''_P(x)}
=\frac{3 T_F}{2T}\left(\tilde{p}
-\frac{1}{\tilde\kappa}\right).
\label{eqn10}
\end{equation}
We note that both compressibility and specific heat depend on the 
second derivatives of the function $F_P(x)$.
\begin{figure}[ht]
\centering
\includegraphics[width=0.8\textwidth]{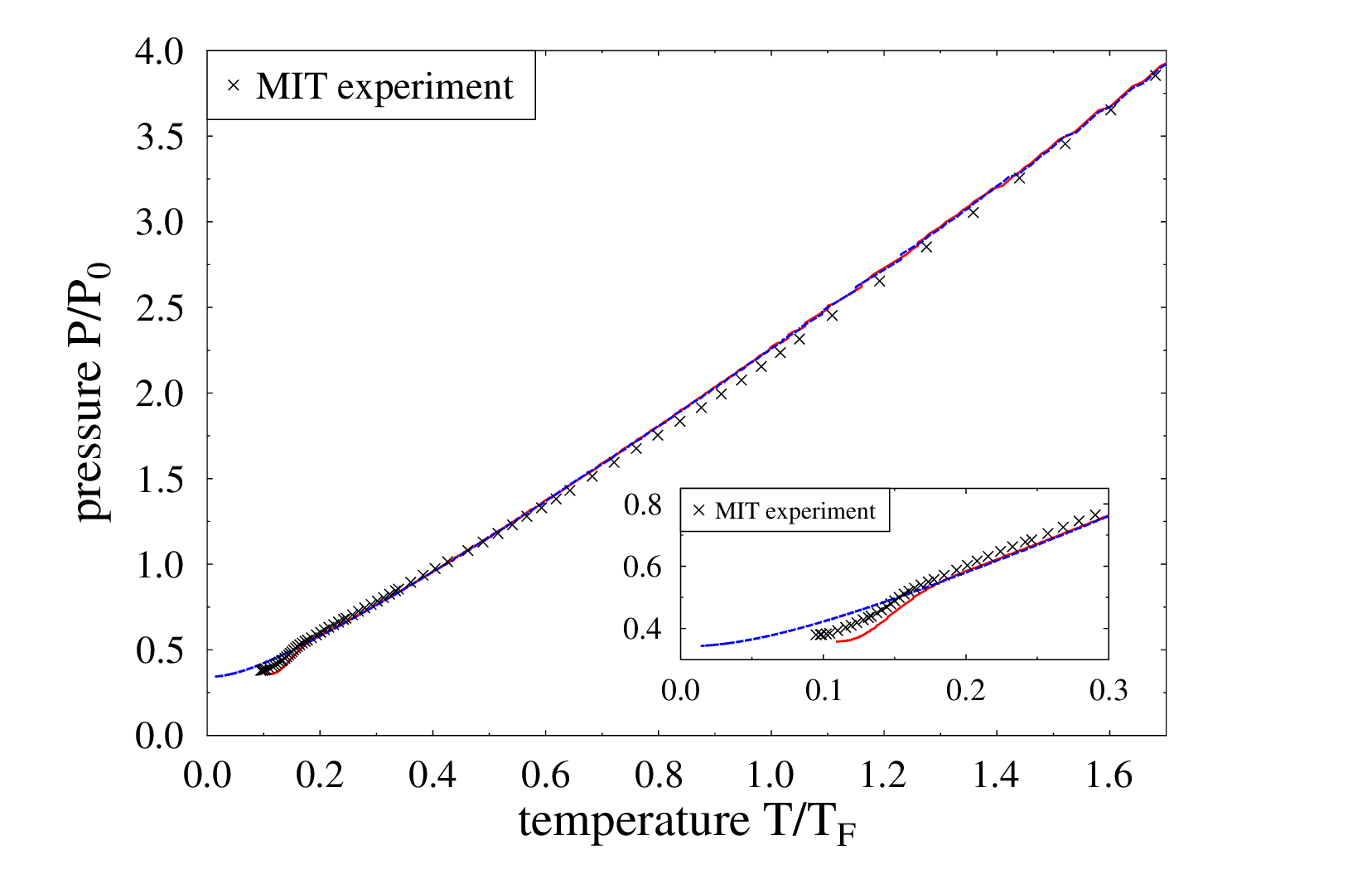}
\vspace*{-0.3cm}
\includegraphics[width=0.8\textwidth]{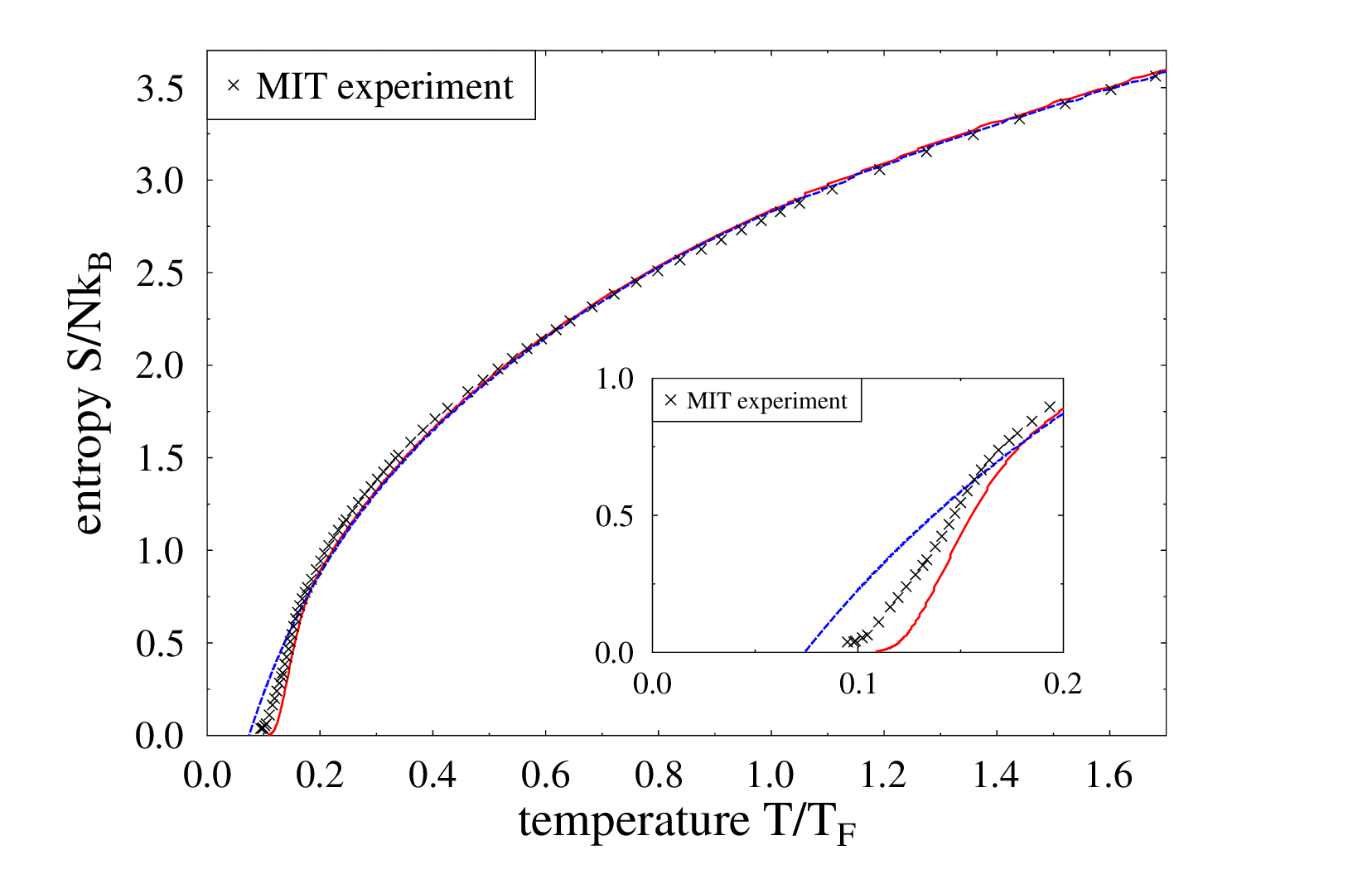}
\vspace*{-0.3cm}
\includegraphics[width=0.8\textwidth]{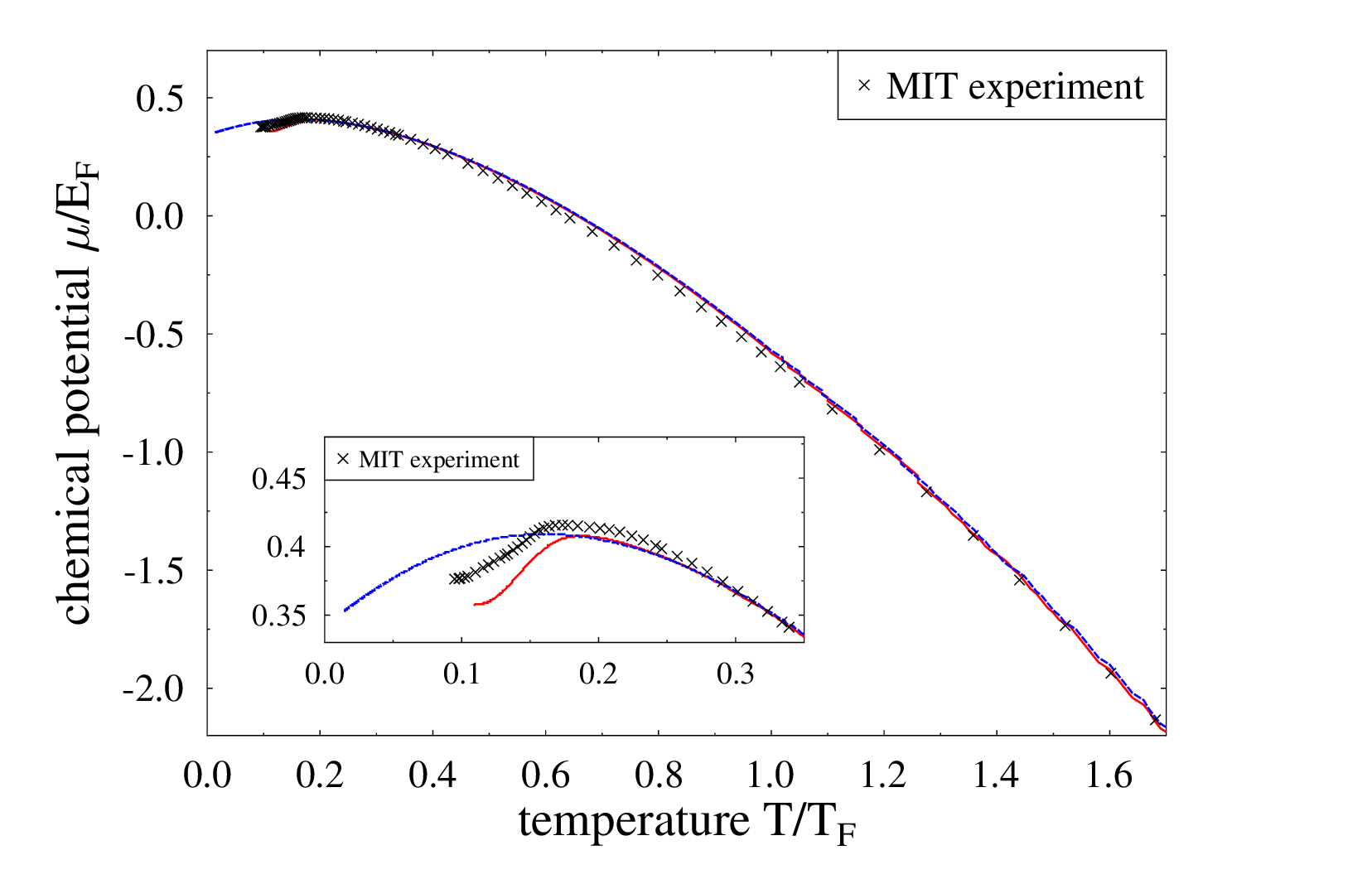}
\vspace*{-0.3cm}
\caption{(Colour online) Normalised pressure (top), entropy (middle)
and chemical potential (bottom) of the unitary Fermi gas as a function 
of temperature. The crosses denote the experimental MIT data \protect\cite{ku}. 
Dashed (blue) and solid (red) lines are as in Fig. \ref{fig1}. The 
inserts give enlarged pictures of the low-temperature domain including 
the phase transition region.}
\label{fig2}
\end{figure}

We now present numerical results for the thermodynamic quantities
for which experimental data are available.
Since no ready-to-use numerical routines for the Fermi-Dirac integrals 
$f_\nu(x)$ could be found, we have calculated them by numerical 
integration of Eq.\ (\ref{eqn2}). This is easily possible to any
desired accuracy for $\nu>1$. For $\nu\leq 1$ we employed the formula 
\cite{pathria} $f_{\nu-1}(x)=f'_\nu(x)$ and used numerical 
differentiation to obtain $f'_\nu(x)$.

In Fig.\ \ref{fig2} we compare our results of normalised pressure (top), 
entropy (middle), and chemical potential (bottom) with the MIT data. 
Our ansatz describes all these data quite well all the way down to the 
critical temperature $T_c=0.16\,T_F$. At high temperatures the results are 
comparable to, if not better than, the virial ansatz discussed in 
Ref.\ \cite{bvm}. Departures are noticed around critical temperature 
$T_c$ (enlarged in the inserts). Like in Fig.\ \ref{fig1}, the results
with and without including $g(x)$ lie on opposite sides of the data below
$T_c$; the solid lines do reproduce the kink seen in the chemical potential
at $T=T_c$. 

We stress again that in Eq.\ (\ref{eqn3}) with $g(x)=0$, only the
contribution $2\times 5f_{5/2}(x)$ is relevant for reproducing the results
at $T=0$. For the energy per particle at $T=0$ it yields 
$E/N=(3/5){\widetilde E}_F$, where ${\widetilde E}_F$ if the Fermi 
energy of the interacting gas. Further, it is easily deduced that 
${\widetilde E}_F/E_F=\xi=(1/5)^{2/3}=0.342$, which is slightly less than 
the experimentally determined value 0.36 of the MIT experiment \cite{ku}.
We emphasise that our fit of $h(x)$ (see Fig.\ref{fig2}) is particularly
sensitive to the linear combination of $f_{5/2}$ and $f_{3/2}$ used in our 
ansatz (\ref{eqn3}), which was primarily chosen to yield a reasonable
high temperature limit.
Any different choice of parameters, or any admixture of Fermi-Dirac integrals 
$f_\nu$ with other orders $\nu$, could not simultaneously yield both these 
desirable large- and zero-temperature limits.

\begin{figure}[htbp]
\centering
\includegraphics[width=0.8\textwidth]{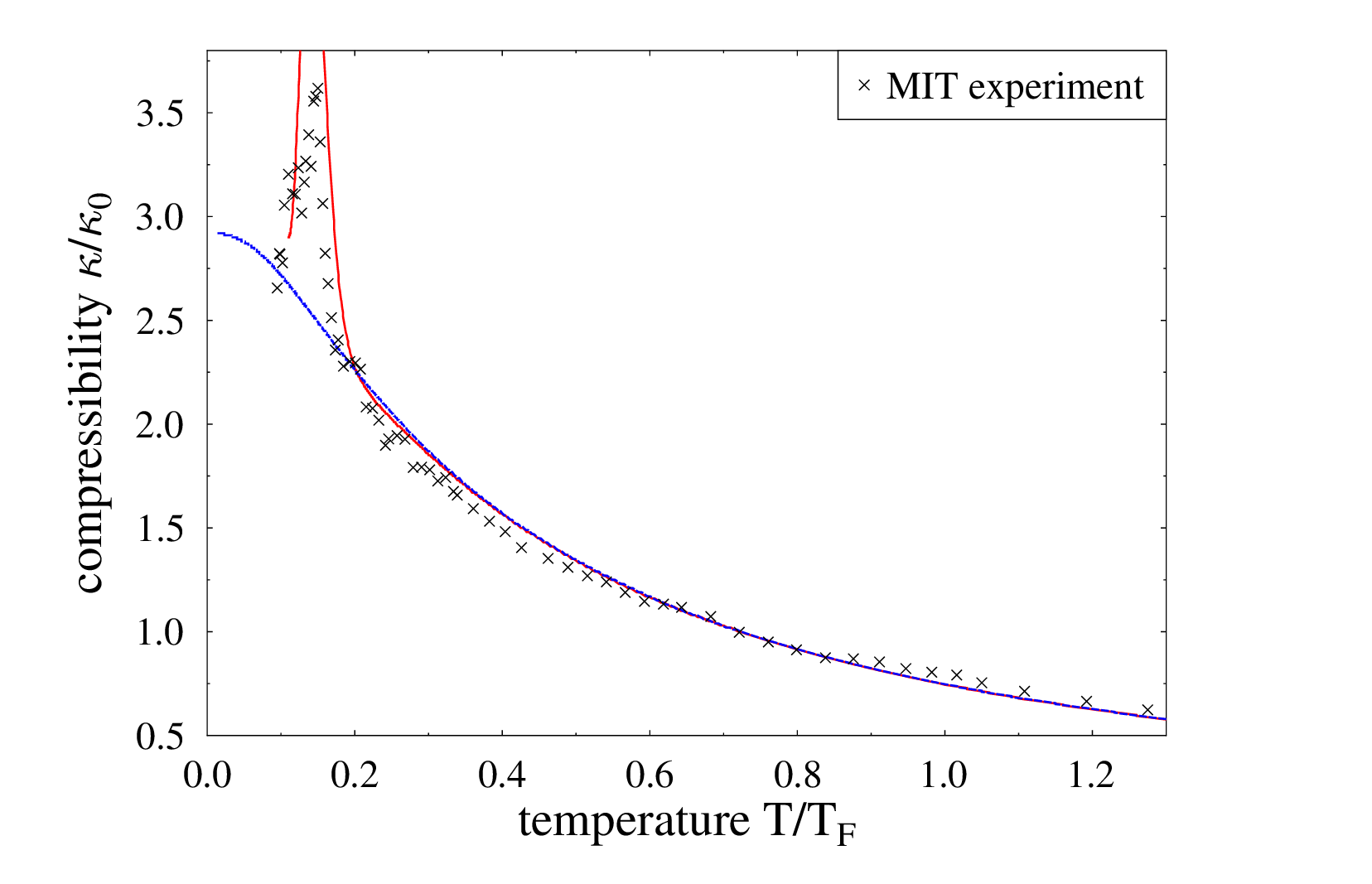}
\vspace*{-0.3cm}
\includegraphics[width=0.8\textwidth]{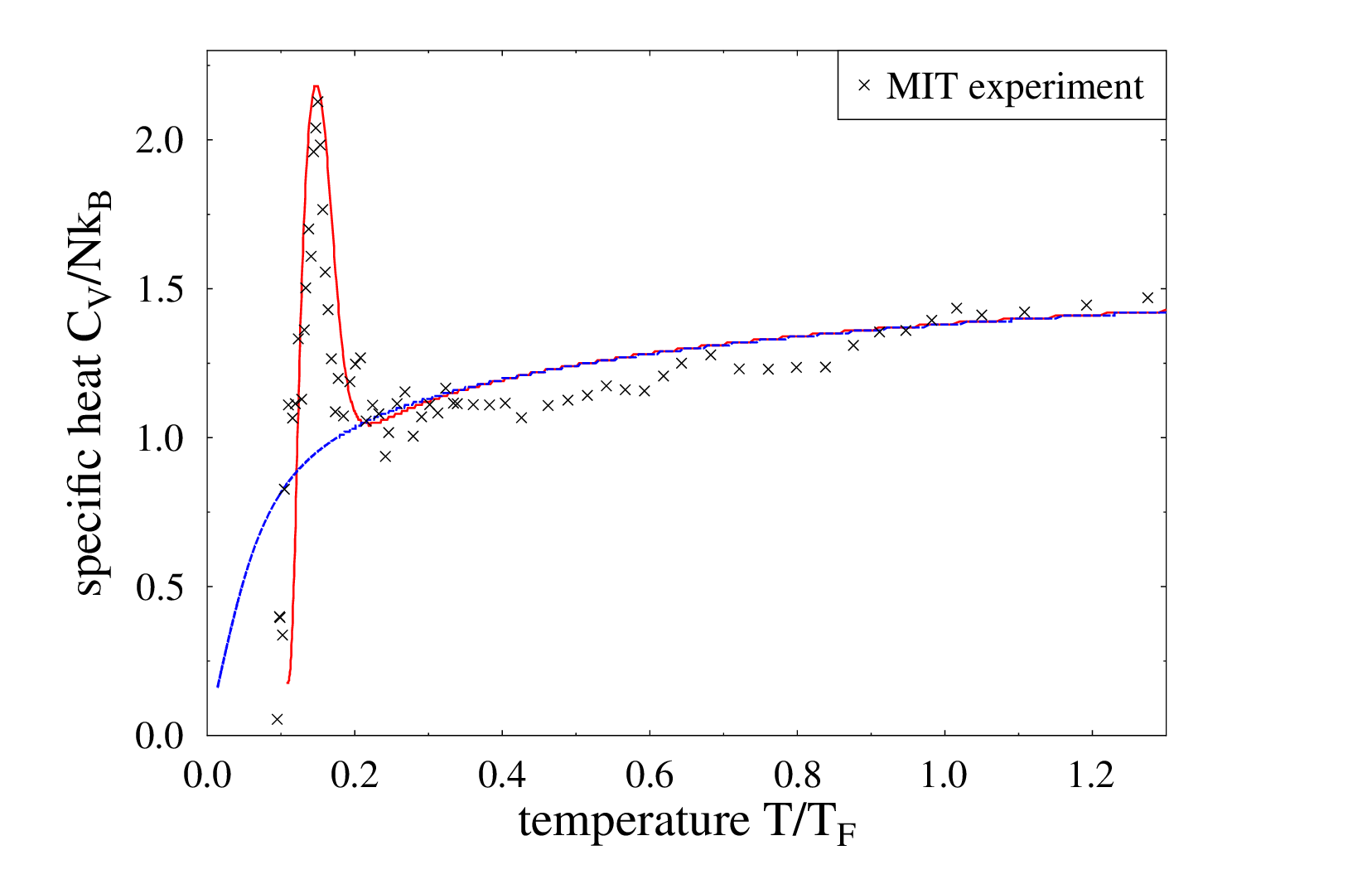}
\vspace*{-0.3cm}
\caption{(Colour online) Normalised compressibility (top) and specific 
heat (bottom) of the unitary Fermi gas as functions of temperature  
(crosses and lines as in the previous figures).}
\label{fig3}
\end{figure}

%\newpage 

\begin{figure}[htbp]
\centering
\includegraphics[width=0.8\textwidth]{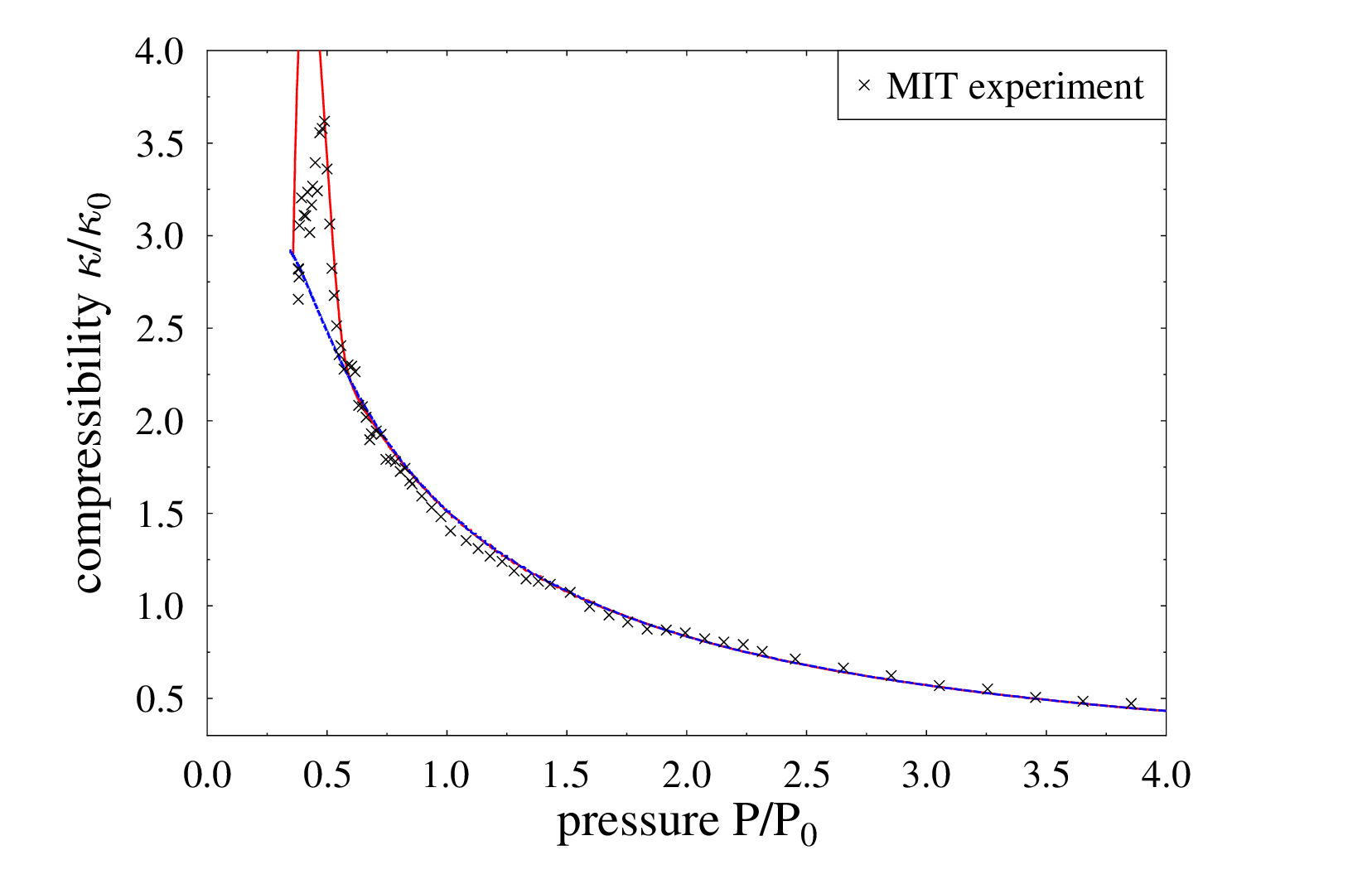}
\vspace*{-0.4cm}
\caption{(Colour online) Normalised compressibility versus normalised
pressure (crosses and lines as in the previous figures).}
\label{fig4}
\end{figure}

In Fig.\ \ref{fig3} we show the results of our calculation for 
compressibility (top) and specific heat (bottom) as functions of 
temperature. The 
resonant term $g(x)$ in $F_P(x)$ here gives an excellent description
of the experimental peaks seen in both quantities. Again, it has to be
cut at $T<T_{min}=0.1\,T_F$, and the result obtained with $g(x)=0$ yields
the correct limits ${\widetilde \kappa}(T$=$0)=1/\xi$ and $C_V(T$=$0)=0$. 
In Fig.\ \ref{fig4} we finally plot compressibility versus normalised
pressure like it was done in Ref.\ \cite{ku}. Note here, in particular,
the excellent agreement with the data up to the highest available 
pressures.

\section{Summary and conclusions}

To summarise, we have introduced a phenomenological function $F_P(x)$, 
given by Eq.\ (\ref{eqn3}) that yields the universal equation of state 
(\ref{eqn1}) of a unitary fermion gas. $F_P(x)$ depends solely on the 
fugacity $z$ (or on $x=\ln z$) and hence is scale independent. It 
consists of two Fermi-Dirac integrals, $f_{5/2}(x)$ and $f_{3/2}(x)$, 
and a resonant term $g(z)$ that corresponds to a pair of zeros of the 
grand partition function in the complex $z$ plane, suitable for 
describing the phase transition observed in the experiments. The 
non-resonant Fermi-Dirac part of $F_P(x)$ is constructed to yield a 
reasonable high temperature limit by imposing the value $\Delta 
b_2=1/\sqrt{2}$ of the second virial coefficient \cite{HoMu}; it 
contains otherwise no adjustable parameter. As a bonus, it also yields 
zero temperature limits that fit the data. The only two parameters $z_c$ 
and $\epsilon$, appearing in Eq.\ (\ref{eqn4}) for the resonant term 
$g(x)$, have been fitted to the critical temperature $T_c=0.16\, T_F$ 
and the width of the phase transition found in the MIT data for specific 
heat and compressibility (see Fig.\ \ref{fig3}). The function $g(z)$ 
diverges for $z\to\infty$, i.e., for $T\to 0$. It was therefore put to 
zero for $z>z_{max}=27$ corresponding to $T<T_{min}=0.1\,T_F$. However, 
the available data seen in the Figures \ref{fig2} - \ref{fig4} are lying 
at $T>T_{min}$, so that we can claim to describe all these data with our 
full ansatz (\ref{eqn3}). The only sizeable deviation is found in Fig.\ 
\ref{fig1} for the quantity $h(x)$ which appears to have been measured 
even below $T_{min}$, and for which our results including $g(x)$ take 
off already above $T_{min}$. Nevertheless, we can claim that our ansatz, 
in spite of its simplicity, describes the overall experimental data 
surprisingly well. To construct it, we have mainly used the universal 
properties of a gas at unitarity, as well as crucial experimental 
observations. 

It is also tempting  to extend the above analysis for trapped fermionic
atoms at unitarity. Following the arguements given above we may write 
$F_P^{(trap)}(x)$ in the form
\begin{equation}                                                   
F_P^{(trap)}(x)=2[f_{4}(x)+4(f_{4}(x)-f_{3}(x))]+g^{(trap)}(x)\,,
\label{trap}
\end{equation}   
which is similar in form to the ansatz given in \ref{eqn3}, with the 
Fermi integrals of the gas replaced by the appropriate Fermi integrals 
for the trap. In the first part there are no new parameters and the 
second virial coefficient is reproduced correctly. This form also 
determines the zero temperature properties of trapped fermionic system 
at unitarity. However, in the absence of data on compressibility and 
specific heat it is not possible to determine the second term $g(x)^{(trap)}$ 
and thus the full form of the thermodynamic potential. Nevertheless, 
the form suggested above may be useful in analysing the results for the 
trap also in future.
 
It would be interesting to see if our $F_P(x)$ can be obtained from a 
microscopic model. We have not succeeded with this, but it is 
hoped that our analysis will trigger future investigations in this 
direction.

\section*{Acknowledgements}

We thank M. Ku and collaborators for sharing their experimental data. 
We acknowledge financial support by NSERC (Canada) and IMSc (India), and 
the hospitality of the Department of Physics and Astronomy, McMaster 
University, and of the Institute of Mathematical Sciences, Chennai.

\end{document}